\documentclass[prl,aps,twocolumn,superscriptaddress,showpacs]{revtex4}
\usepackage{amsmath,amssymb,graphicx}
\usepackage{pxfonts}
\usepackage{graphicx}
\usepackage{color}
\usepackage{float}
\begin{document}

\title{Split-gate device for indirect excitons}

\author{C.~J.~Dorow} \email{cdorow@physics.ucsd.edu} \author{J.~R.~Leonard} \author{M.~M.~Fogler} \author{L.~V.~Butov} \affiliation{Department of Physics, University of California at San Diego, La Jolla, CA 92093, USA}
\author{K.~W.~West} \author{L.~N.~Pfeiffer} \affiliation{Department of Electrical Engineering, Princeton University, Princeton, New Jersey 08544, USA}

\date{\today}

\begin{abstract}
We present a concept and experimental proof of principle for split-gate devices for indirect excitons (IXs). The split-gate forms a narrow channel, a point contact, for IX current. Control of IX flow through the split-gate with both gate voltage and excitation power is demonstrated.
\end{abstract}

\maketitle

Split-gates can be utilized for creating and controlling narrow channels (quantum point contacts) for electrons in mesoscopic electronic devices. Studies of electronic split-gate devices have led to a number of findings including electron focusing~\cite{Sharvin1965, Sharvin1965a, Tsoi1974, Houten1988}, conductance quantization~\cite{Wees1988, Wharam1988}, electron beam collimation~\cite{Molenkamp1990, Eriksson1996, Crook2000}, and electron flow branching~\cite{Topinka2000, Topinka2001}.

In this work, we present a concept and proof-of-principle experiments with split-gate devices for indirect excitons (IXs). An IX is a bound pair of an electron and a hole in spatially separated layers, which can be realized in coupled quantum well (CQW) structures (Fig.~1a). Due to several advantageous properties, IXs form a system that can be used to explore transport of cold bosons through split-gate devices, providing a counterpart to the many transport studies of cold fermions through electronic split-gate devices. These properties include: (i) IXs have built-in dipole moments $ed$, allowing the control of IX energy by voltage, where the IX energy shifts as $\delta E = - edF_z$ ($d$ is the separation between the electron and hole layers, $F_z$ is voltage controllable electric field in the structure growth direction). Various in-plane potential landscapes formed by voltage for IXs were studied in earlier works, including excitonic ramps~\cite{Hagn1995, Gartner2006, Dorow2016}, static~\cite{Zimmermann1997, Zimmermann1998, Hammack2006, Remeika2012, Remeika2015} and moving~\cite{Winbow2011} lattices, traps~\cite{Huber1998, Gorbunov2004, High2009nl, Schinner2013, Shilo2013, Mazuz-Harpaz2017}, and transistors~\cite{Andreakou2014}. (ii) Long IX lifetimes allow them to travel long distances in mesoscopic devices before recombination~\cite{Hagn1995, Gartner2006, Dorow2016, Remeika2012, Remeika2015, Winbow2011, Andreakou2014}. (iii) Long IX lifetimes also allow effective IX thermalization with the crystal lattice~\cite{Butov2001}, giving the opportunity to study IX transport through mesoscopic devices in the quantum regime below the temperature of quantum degeneracy. 

\begin{figure}[htbp]
\includegraphics[width=\linewidth]{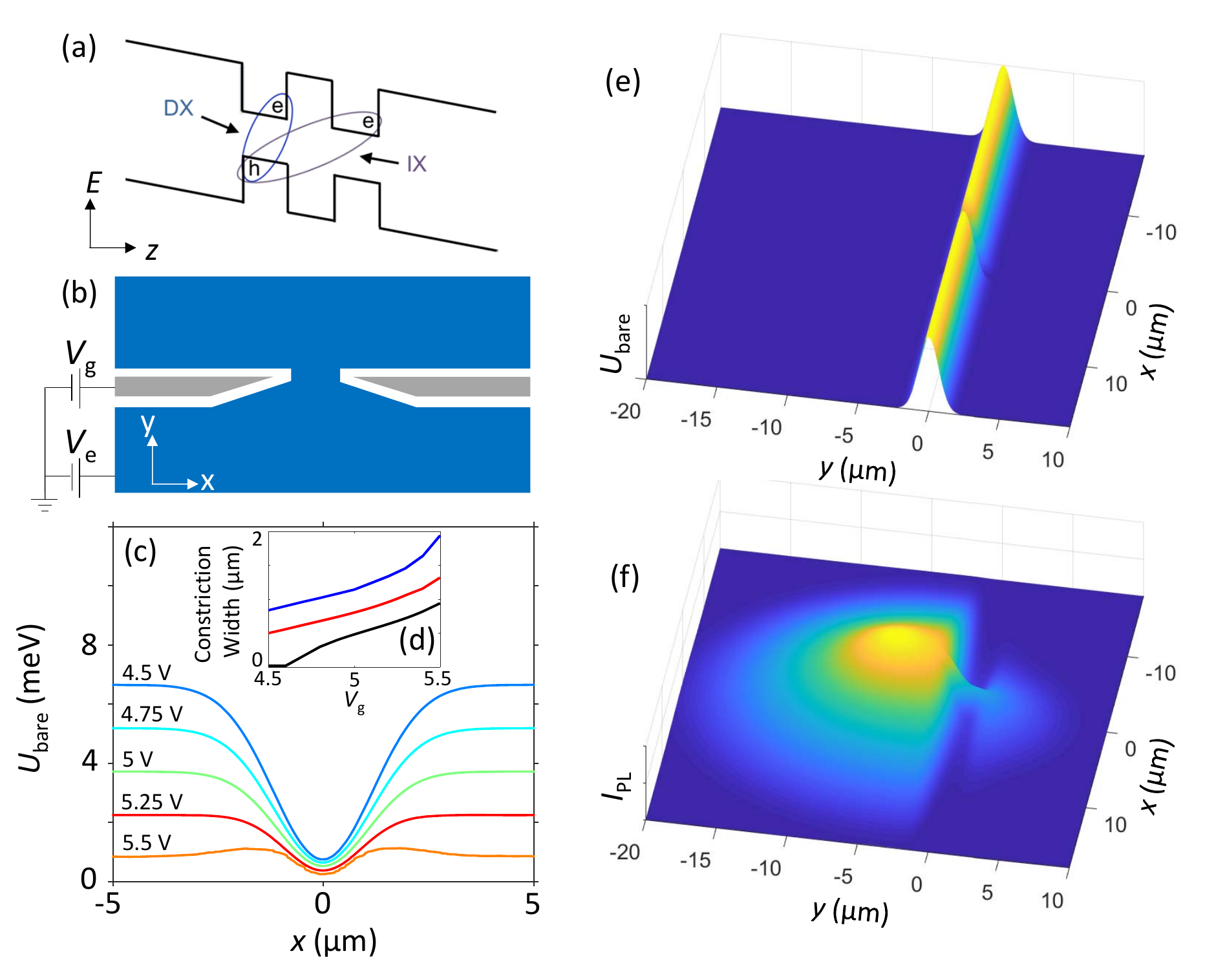}
\caption{(a) CQW band diagram. The ovals indicate a direct exciton (DX) and an indirect exciton (IX) composed of an electron ($e$) and a hole ($h$). (b) Schematic of the excitonic split-gate device. (c) Simulated bare constriction energy profiles for different split-gate voltages $V_{\rm g}$ for fixed voltage on large electrode, $V_{\rm e} = 5.5$~V. (d) The corresponding constriction width $w$ for IXs with different energies $E = 0.5$ (black), 0.7 (red) and 1 meV (blue) vs $V_{\rm g}$. (e) Simulated IX potential energy in the bare split-gate device. (f) Simulated IX PL in potential of (e).}
\end{figure}

\begin{figure*}[htbp]
\includegraphics[width=13.5cm]{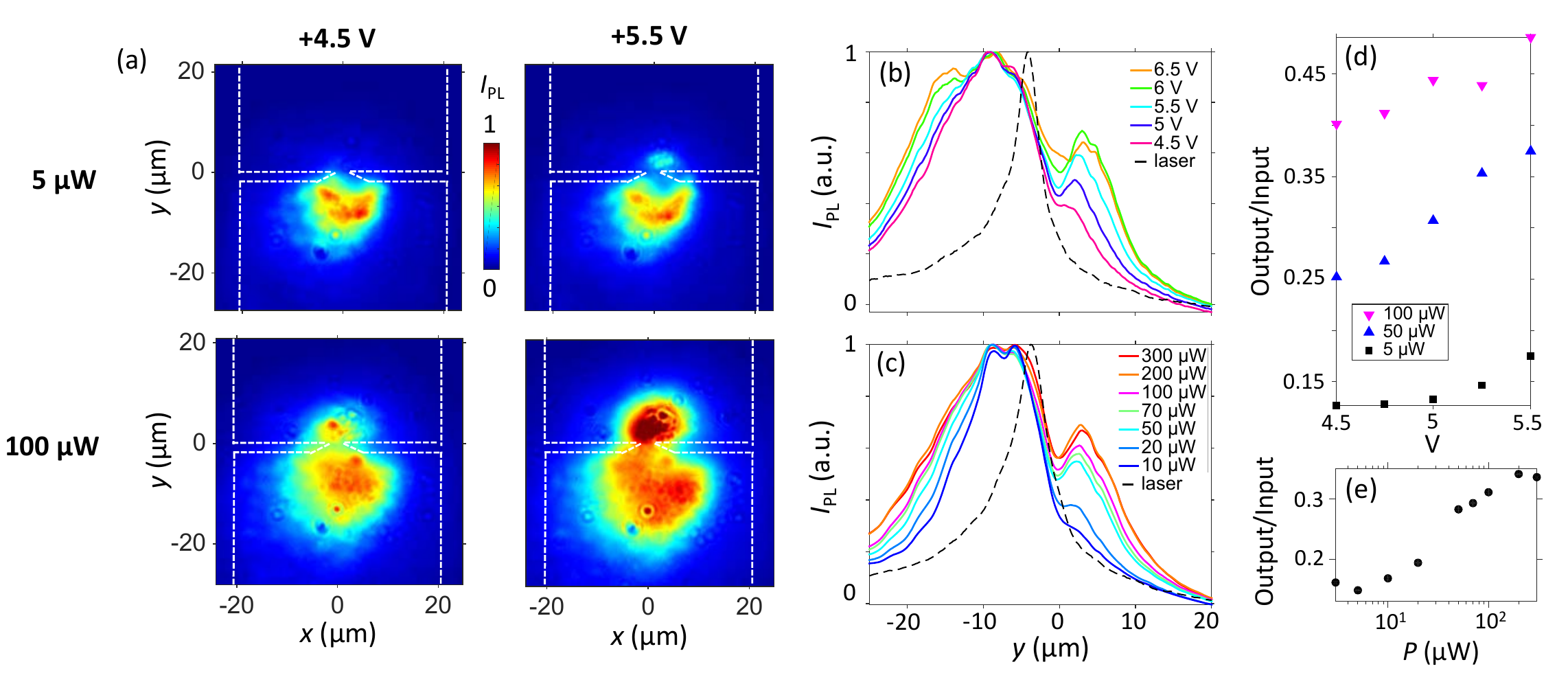}
\caption{(a) IX PL emission for two split-gate voltages $V_{\rm g} = 4.5$ and 5.5~V and two laser excitation powers $P = 5$ and 100~$\mu$W. Contour of the large electrode is shown with dashed lines. (b) Normalized IX PL intensity integrated over $x$ vs $V_{\rm g}$. $P = 50$~$\mu$W. (c) Normalized IX PL intensity integrated over $x$ vs $P$. $V_{\rm g} = 4.5$~V. (d) IX flux through constriction vs $V_{\rm g}$. $P = 5$, 50, and 100~$\mu$W. (e) IX flux through constriction vs $P$. $V_{\rm g} = 4.5$~V. Large electrode is held at $V_{\rm e}= 5.5$~V for all data.}
\end{figure*}

An excitonic split-gate device is formed by two electrodes: a large electrode (shown in blue (dark) in Fig.~1b) and a split-gate electrode (shown in gray in Fig.~1b). Voltage $V_{\rm e}$ on the large electrode realizes the indirect regime in which IXs form the lowest energy state with energies below the energy of spatially direct excitons (DXs), which are also shown in Fig.~1a. Gate voltage $V_{\rm g}$ on the split-gate electrode creates a narrow channel for IXs (Fig.~1c-f). The design of the IX split-gate device is similar to the design of electronic split-gate devices in semiconductor structures~\cite{Houten1988, Wees1988, Wharam1988, Molenkamp1990, Eriksson1996, Crook2000, Topinka2000, Topinka2001}. A difference is in the presence of the large electrode, which is needed to implement the indirect regime for IX devices. This large electrode is separated from the split-gate electrode by a narrow opening. We note that in some IX CQW samples, e.g. in GaAs/AlAs CQW samples studied in~\cite{Hagn1995, Huber1998, Hammack2006}, the indirect regime is realized at $V_{\rm e} = 0$. Therefore, these samples don't require a large electrode and, in turn, the narrow opening between it and the split-gate electrode, making the device design simpler.

The electric field $F_z(x,y)$ and resulting IX potential energy $U_{\rm bare}(x,y) = - edF_z(x,y)$ for the bare, unscreened, split-gate device were modeled by numerically solving Poisson’s equation (note that the split-gate potential landscape and, in turn, $w$ are affected by IX screening, this is discussed below). Cross-sections of IX energy profiles at the split-gate position are shown in Fig.~1c for different $V_{\rm g}$. The IX energy is given relative to the energy of IXs far from the split-gate electrode, as determined by $V_{\rm e}$ on the large electrode.

The channel width $w$ for transport of IXs with energy $E$ (relative to the IX energy in a bare device away from the split-gate) is controlled by $V_{\rm g}$. This is illustrated in Fig.~1d for a bare potential for several $E$ values. The control of split-gate channel by voltage provides control of IX current passing through. At low temperatures, $E$ is determined by IX interaction. IXs are oriented dipoles and interact repulsively with the interaction energy on the order of meV for typical IX densities~\cite{Ivanov2002, Remeika2015}. The IX interaction energy for IX split-gate devices is analogous to the electron Fermi energy for electronic split-gate devices.

Figure~1e,f presents 3D images illustrating IX transport through a split-gate. IXs are generated by laser excitation on one side of the split-gate device (laser excitation centered around $y = - 5$~$\mu$m is shown in Fig. 1f) and travel to the other side through the split-gate. This geometry corresponds to the experiments described below.

In the CQW structure grown by molecular beam epitaxy, an $n^+$-GaAs layer with n$_{\rm Si} = 10^{18}$~cm$^{-3}$ serves as a homogeneous bottom electrode. Two 8~nm GaAs QWs are separated by a 4~nm Al$_{0.33}$Ga$_{0.67}$As barrier and positioned 0.1~$\mu$m above the $n^+$-GaAs layer within an undoped 1~$\mu$m thick Al$_{0.33}$Ga$_{0.67}$As layer. The QWs are positioned close to the homogeneous bottom electrode to suppress the in-plane electric field~\cite{Hammack2006}, which could otherwise lead to IX dissociation~\cite{Zimmermann1997}. The top semitransparent electrodes are fabricated by applying 2~nm Ti and 7~nm Pt.

IXs were generated by a 633 nm HeNe laser focused to a spot with full width half maximum $\sim 5$~$\mu$m. Exciton photoluminescence (PL) was measured by a spectrometer and a liquid nitrogen cooled charge coupled device camera (CCD). The spatial $x$--$y$ IX PL pattern was measured by the CCD after spectral selection by an $800 \pm 5$~nm interference filter chosen to match the IX energy. As a result, the low-energy bulk emission, higher-energy DX emission, and scattered laser light were effectively removed and the IX PL images were visualized. Experiments were performed in an optical helium cryostat at bath temperature $T_{\rm bath} = 1.7$~K.

Experimental proof of principle for the IX split-gate device is shown in Fig.~2. IXs are photogenerated on one side of the split-gate (laser is positioned at (0,-4~$\mu$m) in Fig.~2) and their transport through the split-gate is detected by measuring the spatial pattern of the IX emission. The rows in Fig.~2a and Figs.~2b and 2d show control of IX transport through the split-gate by gate voltage $V_{\rm g}$ for a fixed laser excitation power $P$. Increasing the channel width $w$ by voltage enhances the IX flux through the split-gate.

\begin{figure}[htbp]
\includegraphics[width=\linewidth]{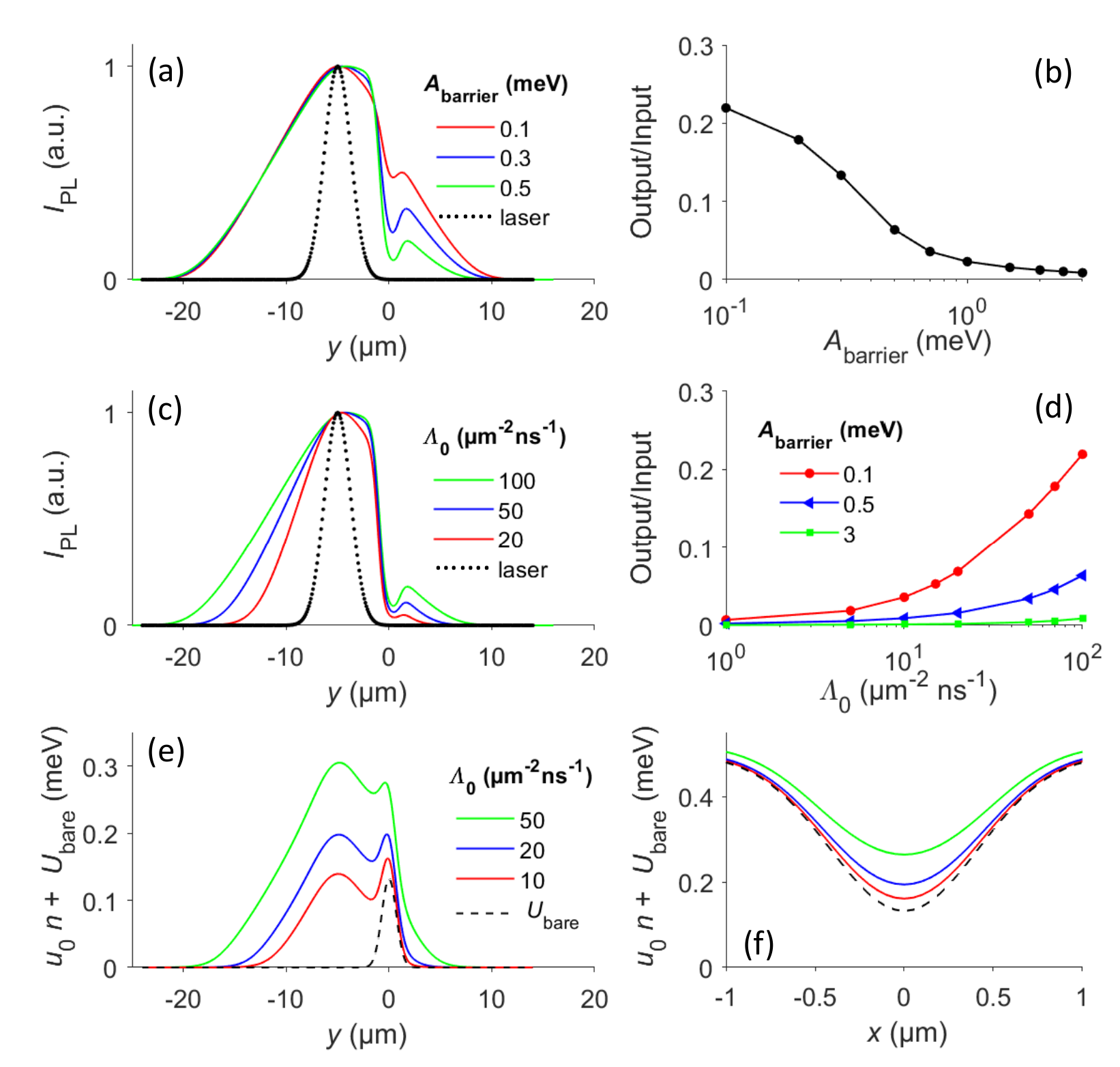}
\caption{Results of simulations. (a) IX PL vs bare barrier height $A_{\rm barrier}$. IX generation rate $\Lambda_0 = 100$~$\mu$m$^{-2}$ns$^{-1}$. IX PL is integrated over $x$. The laser profile is indicated with a dashed line. (b) IX flux through constriction vs $A_{\rm barrier}$. $\Lambda_0 = 100$~$\mu$m$^{-2}$ns$^{-1}$. (c) IX PL vs $\Lambda_0$. $A_{\rm barrier} = 0.5$~meV. IX PL is integrated over $x$. (d) IX flux through constriction vs $\Lambda_0$. $A_{\rm barrier} = 0.1$, 0.5, and 3~meV. (e,f) IX energy (e) vs $y$ at $x=0$ and (f) vs $x$ at $y=0$ for several $\Lambda_0$. The bare barrier profile $U_{\rm bare}$ is shown by dashed lines. $A_{\rm barrier} = 0.5$~meV. For all simulations, $T_{\rm bath} = 1.7$~K.
}
\end{figure}

The columns in Fig.~2a and Figs.~2c and 2e show control of IX transport through the split-gate by laser excitation power $P$ for a fixed gate voltage $V_{\rm g}$. Increasing IX density by excitation power enhances the IX flux through the split-gate. This is described in terms of screening below.

We simulated IX transport through the split-gate in the diffusive regime corresponding to the proof of principle experiments shown in Fig. 2. This regime is characterized by the mean free path small compared to the device dimensions. The following nonlinear partial differential equation was used to model IX transport through the split-gate:
\begin {equation}
\nabla \cdot \left[D \nabla n + \mu n \nabla(u_{\rm 0}n + U_{\rm bare}) \right] + \Lambda - \frac{n}{\tau} = 0.
\label{transport}
\end {equation}
The first term in square brackets in Eq.~1 accounts for IX diffusion, $n$ is the IX density, $D$ the IX diffusion coefficient. The second term accounts for IX drift due to the dipole-dipole IX interaction, which is approximated by the plate capacitor formula $u_{\rm 0}n = \frac{4 \pi e^2 d}{\varepsilon}n$, $\varepsilon$ is the dielectric constant~\cite{Ivanov2002}, and due to the split-gate potential $U_{\rm bare}(x,y) = - e d F_{\rm z}(x,y)$. The IX mobility $\mu$ is given by the Einstein relation $\mu = D/(k_{\rm B} T)$. The effect of in-plane disorder intrinsic to QWs is approximated using a ``thermionic model'' for the diffusion coefficient, $D = D^{(0)}{\rm exp}\left(-U_{\rm dis}/(k_{\rm B}T + u_{\rm 0}n)\right)$~\cite{Ivanov2002}. $D^{(0)}$ is the diffusion coefficient in the absence of QW disorder and $U_{\rm dis}/2$ is the amplitude of the disorder potential. The temperature of IXs $T$ is approximated as $T = T_{\rm bath}$. The non-resonant photoexcitation causes heating of the IX gas by a few Kelvin. However, the hot IXs cool to the lattice temperature within few microns of the excitation spot~\cite{Hammack2009} justifying the approximation. The last two terms in Eq.~1 account for the creation and decay of IXs. $\Lambda_0(x,y)$ is the IX generation rate and $\tau$ is the IX lifetime.

Simulations (Fig.~3) qualitatively reproduce both the control of IX transport through the split-gate by voltage $V_{\rm g}$ (compare Fig.~2b with Fig.~3a and Fig.~2d with Fig.~3b) and by excitation power $P$ (compare Fig.~2c with Fig.~3c and Fig.~2e with Fig.~3d). The data are discussed below.

Increasing absolute value of gate voltage $V_{\rm g}$ increases the channel width $w$ at the IX energy and also reduces the barrier height (Fig.~1c). As a result, IX transport through the split-gate is controlled by gate voltage $V_{\rm g}$ (rows in Fig.~2a and Figs.~2b, 2d, 3a, and 3b). The dependence on voltage in simulations is presented by the dependence on the height of the bare barrier away from the channel $A_{\rm barrier}$ (e.g. $A_{\rm barrier} = 3$~meV corresponds to $V_{\rm g} = 5.1$~V, see Fig.~1c).

Increasing IX excitation power $P$ increases IX density $n$. This causes screening of the split-gate potential by IXs, increasing the channel width and reducing the barrier for IX transport through the split-gate (Figs.~3e and 3f). Increasing $n$ also causes screening of disorder as IXs interact repulsively. Screening of the split-gate potential and disorder increases IX transport through the split-gate (columns in Fig.~2a and Figs.~2c, 2e, 3c, and 3d). 

An interesting regime for electron transport through electronic split-gates is the regime of quantum ballistic transport, where the mean free path and Fermi wavelength exceed the device dimensions. This regime is realized for electronic split-gate devices in high-quality semiconductor structures at low temperatures ~\cite{Houten1988, Wees1988, Wharam1988, Molenkamp1990, Eriksson1996, Crook2000, Topinka2000, Topinka2001}.

For excitonic devices, at low temperatures ($T \sim 0.1$~K, achievable in dilution refrigerators~\cite{Butov2001, High2012}), the IX coherence length in a coherent IX gas in high-quality CQW semiconductor structures reaches $\sim 10$~$\mu$m~\cite{High2012}, exceeding the dimensions of the split-gate channel (Fig.~1), IX interparticle separation ($\sim 0.1$~$\mu$m for typical IX density $10^{10}$~cm$^{-2}$), and IX thermal de Broglie wavelength ($\sim 0.5$~$\mu$m for IX temperature $T = 0.1$~K). This indicates the feasibility of the realization of IX quantum ballistic transport through excitonic split-gate devices at low temperatures. The realization of this regime forms the subject for future works.

We note that excitonic split-gate devices allow imaging IX current paths after spatially localized IX injection. Therefore, besides giving the opportunity to extend studies of narrow-channel phenomena in fermions~\cite{Sharvin1965, Sharvin1965a, Tsoi1974, Houten1988, Wees1988, Wharam1988, Molenkamp1990, Eriksson1996, Crook2000, Topinka2000, Topinka2001} to bosons, IX split-gate devices can also be used as a tool to probe directional effects in transport of composite bosons, including the predicted exciton Hall effect~\cite{Dzyubenko1984} and exciton spin Hall effect~\cite{Kavokin2005}. Excitonic split-gate devices can also be used for studying transport of composite particles through narrow channels~\cite{Grasselli2016}.

In conclusion, we presented a concept and experimental proof of principle for split-gate devices for indirect excitons.

This work was supported by NSF Grant No.~1640173 and NERC, a subsidiary of SRC, through the SRC-NRI Center for Excitonic Devices. C.~J.~D. was supported by the NSF Graduate Research Fellowship Program under Grant No.~DGE-1144086. This work used the Extreme Science and Engineering Discovery Environment (XSEDE), which is supported by NSF grant ACI-1548562, and XSEDE computer Comet at the San Diego Supercomputer Center through allocation TG-ASC150024.

\end{document}